\documentclass[a4paper,english,preprint]{revtex4}
\usepackage[T1]{fontenc}
\usepackage[latin9]{inputenc}
\usepackage[active]{srcltx}
\usepackage{color}
\usepackage{babel}
\usepackage{units}
\usepackage{amsmath}
\usepackage{amssymb}
\usepackage{graphicx}
\usepackage{esint}
\usepackage[unicode=true,
 bookmarks=true,bookmarksnumbered=true,bookmarksopen=false,
 breaklinks=false,pdfborder={0 0 0},backref=false,colorlinks=true]
 {hyperref}
\hypersetup{pdftitle={Spin-dependent part of pd  interaction cross section and Nijmegen potential},
 pdfauthor={S. G. Salnikov},
 pdfkeywords={polarization, antiprotons, Nijmegen, Glauber, deuteron},
 citecolor=blue,pdfstartview=FitH}
\usepackage{breakurl}

\makeatletter

\@ifundefined{textcolor}{}
{%
 \definecolor{BLACK}{gray}{0}
 \definecolor{WHITE}{gray}{1}
 \definecolor{RED}{rgb}{1,0,0}
 \definecolor{GREEN}{rgb}{0,1,0}
 \definecolor{BLUE}{rgb}{0,0,1}
 \definecolor{CYAN}{cmyk}{1,0,0,0}
 \definecolor{MAGENTA}{cmyk}{0,1,0,0}
 \definecolor{YELLOW}{cmyk}{0,0,1,0}
 }

\makeatother

\begin{document}

\title{Spin-dependent part of $\bar{p}d$ interaction cross section and
Nijmegen potential.}

\author{S.~G.~Salnikov}

\email{salsergey@gmail.com}

\affiliation{Budker Institute of Nuclear Physics, 630090 Novosibirsk, Russia}

\date{\today}
\begin{abstract}
Low energy $\bar{p}d$ interaction is considered taking into account
the polarization of both particles. The corresponding cross sections
are obtained using the Nijmegen nucleon-antinucleon optical potential
with shadowing effects taken into account. Double-scattering effects
are calculated within the Glauber approach and found to be about $10\div20\%$.
The cross sections are applied to the analysis of the polarization
buildup which is due to the interaction of stored antiprotons with
a polarized target. It is shown that, at realistic parameters of a
storage ring and a target, the filtering mechanism may provide a noticeable
polarization in a time comparable with the beam lifetime. The energy
dependence of the polarization rate for deuterium target is similar
to that for hydrogen one. However, the time of polarization for deuterium
is much smaller than that for hydrogen.
\end{abstract}



\maketitle

\section{Introduction}

An extensive research program with polarized antiprotons has been
proposed recently by the PAX Collaboration~\cite{PAX05}. This program
initiated a discussion of various methods to polarize stored antiprotons.
One of the methods being considered is to use multiple scattering
off a polarized target. If all particles remain in the beam (scattering
angle is smaller than the acceptance angle $\theta_{\textrm{acc}}$),
only spin flip can lead to polarization buildup, as was shown in Refs.~\cite{MilStr05,NikPav06}.
However, spin\nobreakdash-flip cross section is negligibly small~\cite{MilStr05,MilSalStr08}.
Hence the most realistic method is spin filtering~\cite{Csonka68}.
This method implements the dependence of the cross section on orientation
of the spins of the particles. Therefore the number of antiprotons
scattered out of the beam after the interaction with a polarized target
depends on their spins, which results in the polarization buildup.
The interaction with atomic electrons can't provide noticeable polarization
because in this case antiprotons will scatter only in small angles
and all antiprotons remain in the beam~\cite{MilStr05}. Thus it
is necessary to study antiproton-nuclear scattering.

At present, theory can't give reliable predictions for $\bar{p}N$~cross
section below~$\unit[1]{GeV}$ and different phenomenological models
are usually used for numerical estimations. As a result, the cross
sections obtained are model-dependent. All models are based on fitting
experimental data for scattering of unpolarized particles. These models
give similar predictions for spin-independent part of the cross sections,
but predictions for spin-dependent parts may differ drastically.

Different nucleon-antinucleon potentials have similar behavior at
large distance (\mbox{$r\gtrsim\unit[1]{fm}$}) because long\nobreakdash-range
potentials are obtained by applying G\nobreakdash-parity transformation
to well\nobreakdash-known nucleon-nucleon potential. The most important
difference between nucleon-antinucleon and nucleon-nucleon scattering
is existence of annihilation channels. A phenomenological description
of annihilation is usually based on an optical potential of the form
\begin{equation}
V_{N\bar{N}}=U_{N\bar{N}}-iW_{N\bar{N}}.
\end{equation}
Imaginary part of this potential describes annihilation into mesons
and is important at small distance. The process of annihilation has
no generally accepted description, and short\nobreakdash-range potentials
in various models are different.

One of the methods to polarize antiprotons being investigated is to
use scattering off a polarized hydrogen target. Spin-dependent parts
of the cross section of $\bar{p}p$~interaction were previously calculated
in Ref.~\cite{DmitMilStr08} using the Paris potential and in Ref.~\cite{DmitMilSal10}
with the help of the Nijmegen potential. Similar calculations were
performed in Ref.~\cite{PAX09} where various forms of Julich potentials
were explored. All models listed above predict a possibility to obtain
a noticeable beam polarization in a reasonable time, but the value
of the polarization degree predicted is essentially different.

Another possibility to polarize stored antiprotons being considered
is to use polarized deuteron target instead of a hydrogen target.
Theoretical investigation of antiproton-deuteron scattering is the
subject present paper is devoted to. We make use of the Nijmegen model
to calculate $\bar{p}N$ scattering amplitudes. In order to calculate
$\bar{p}d$ cross sections we utilize the Glauber theory~\cite{FrancoGlauber66,FrancoGlauber69}.
We believe that the Glauber approach has sufficient precision for
the description of $\bar{p}d$ scattering in the energy region concerned.
The Figures confirming this statement are presented in Sec.~\ref{sec:Results}.
In the present paper we show our predictions for the spin-dependent
parts of $\bar{p}d$  cross sections along with the expected antiproton
beam polarization degree. The comparison with the predictions from
Ref.~\cite{UzikHaiden09} based on the Julich models are also shown
below.

\section{Method of calculation}

Our method of calculation is similar to that described in Ref.~\cite{UzikHaiden09}.
We make use of the Glauber theory to describe scattering by a deuteron.
In the present paper we give the formulas for the standard Glauber
theory~\cite{FrancoGlauber66} which doesn't include the D-wave contribution
in the deuteron wave function and the spin dependence of $\bar{p}N$
scattering amplitudes. The modification of this theory taking these
factors into account for the case of $pd$ scattering can be found
in Ref.~\cite{PlatKuk10}. Within the standard Glauber theory the
amplitudes for elastic and breakup scattering are given by the following
matrix elements
\begin{equation}
F_{fi}^{\bar{p}d}\left(\boldsymbol{q}\right)=\left\langle f\right|F^{\bar{p}d}\left(\boldsymbol{q},\boldsymbol{s}\right)\left|i\right\rangle \label{eq:Amplitude}
\end{equation}
between initial $\left|i\right\rangle $ and final $\left|f\right\rangle $
states of the two\nobreakdash-nucleon system. Here the transition
operator is
\begin{equation}
F^{\bar{p}d}(\boldsymbol{q},\boldsymbol{s})=e^{\frac{1}{2}i\boldsymbol{q}\cdot\boldsymbol{s}}f_{\overline{p}p}(\boldsymbol{q})+e^{-\frac{1}{2}i\boldsymbol{q}\cdot\boldsymbol{s}}f_{\overline{p}n}(\boldsymbol{q})+\frac{i}{2\pi k_{\bar{p}d}}\int e^{i\boldsymbol{q}'\cdot\boldsymbol{s}}f_{\overline{p}p}({\textstyle \frac{1}{2}}\boldsymbol{q}-\boldsymbol{q}')f_{\overline{p}n}({\textstyle \frac{1}{2}}\boldsymbol{q}+\boldsymbol{q}')d^{2}\boldsymbol{q}',\label{eq:GlauberApproximation}
\end{equation}
where $\boldsymbol{q}$~is the momentum transfer, $\boldsymbol{s}$~is
the impact parameter (the transverse component of $\boldsymbol{r}$),
$f_{\overline{p}N}(\boldsymbol{q})$~are antiproton-nucleon elastic
scattering amplitudes and $k_{\bar{p}d}=\sqrt{m_{N}T_{\mathrm{lab}}/2}$~is
the antiproton momentum, $m_{N}$~being the nucleon mass and $T_{\mathrm{lab}}$~
being the antiproton kinetic energy in the laboratory frame. Note
that the antiproton momentum and antiproton-nucleon scattering amplitudes
should be calculated in the same reference system. Using Eqs. \eqref{eq:Amplitude}
and \eqref{eq:GlauberApproximation} one obtains the following equation
for the elastic antiproton-deuteron scattering amplitude:
\begin{equation}
F_{ii}^{\bar{p}d}(\boldsymbol{q})=S\left(\boldsymbol{q}\right)f_{\overline{p}p}(\boldsymbol{q})+S\left(-\boldsymbol{q}\right)f_{\overline{p}n}(\boldsymbol{q})+\frac{i}{2\pi k_{\bar{p}d}}\int S\left(\boldsymbol{q}'\right)f_{\overline{p}p}({\textstyle \frac{1}{2}}\boldsymbol{q}-\boldsymbol{q}')f_{\overline{p}n}({\textstyle \frac{1}{2}}\boldsymbol{q}+\boldsymbol{q}')d^{2}\boldsymbol{q}'.\label{eq:ElasticAmplitude}
\end{equation}
Note that the latter formula involves only the elastic deuteron form
factor $S\left(\boldsymbol{q}\right)$ and amplitudes of $\bar{p}N$
scattering. Elastic ($\bar{p}d\to\bar{p}d$) differential cross section
is given by
\begin{equation}
\left(\frac{d\sigma}{d\Omega}\right)_{\mathrm{el}}=\left|F_{ii}^{\bar{p}d}(\boldsymbol{q})\right|^{2}.
\end{equation}
If one neglects the energy difference of various final states then
the sum of elastic plus inelastic ($\bar{p}d\to\bar{p}\, pn$) cross
sections can be calculated in the following way:
\begin{multline}
\left(\frac{d\sigma}{d\Omega}\right)_{\mathrm{sc}}=\sum_{f}\left|F_{fi}^{\bar{p}d}(\boldsymbol{q})\right|^{2}=\left\langle i\right|\left|F^{\bar{p}d}\left(\boldsymbol{q},\boldsymbol{s}\right)\right|^{2}\left|i\right\rangle =\left|f_{\overline{p}p}(\boldsymbol{q})\right|^{2}+\left|f_{\overline{p}n}(\boldsymbol{q})\right|^{2}+2S(\boldsymbol{q})\mathrm{Re}\left[f_{\bar{p}n}(\boldsymbol{q})f_{\bar{p}p}^{*}(\boldsymbol{q})\right]\\
-\frac{1}{\pi k_{\bar{p}d}}\mathrm{Im}\left[f_{\bar{p}n}^{*}(\boldsymbol{q})\int S\left(\boldsymbol{q}'-{\textstyle \frac{1}{2}}\boldsymbol{q}\right)f_{\overline{p}p}({\textstyle \frac{1}{2}}\boldsymbol{q}-\boldsymbol{q}')f_{\overline{p}n}({\textstyle \frac{1}{2}}\boldsymbol{q}+\boldsymbol{q}')d^{2}\boldsymbol{q}'\right]\\
-\frac{1}{\pi k_{\bar{p}d}}\mathrm{Im}\left[f_{\bar{p}p}^{*}(\boldsymbol{q})\int S\left(\boldsymbol{q}'+{\textstyle \frac{1}{2}}\boldsymbol{q}\right)f_{\overline{p}p}({\textstyle \frac{1}{2}}\boldsymbol{q}-\boldsymbol{q}')f_{\overline{p}n}({\textstyle \frac{1}{2}}\boldsymbol{q}+\boldsymbol{q}')d^{2}\boldsymbol{q}'\right]\\
+\frac{1}{\left(2\pi k_{\bar{p}d}\right)^{2}}\int d^{3}\boldsymbol{r}\left|\phi(\boldsymbol{r})\right|^{2}\left|\int e^{i\boldsymbol{q}'\cdot\boldsymbol{s}}f_{\overline{p}p}({\textstyle \frac{1}{2}}\boldsymbol{q}-\boldsymbol{q}')f_{\overline{p}n}({\textstyle \frac{1}{2}}\boldsymbol{q}+\boldsymbol{q}')d^{2}\boldsymbol{q}'\right|^{2}.
\end{multline}

The amplitudes of antinucleon-nucleon scattering were calculated with
the help of the Nijmegen antinucleon-nucleon optical potential~\cite{TimmRijSw94,NagRijSw78}
in the same way as in our previous work~\cite{DmitMilSal10}.

The total spin-dependent $\bar{p}d$ cross section can be written
in the form~\cite{UzikHaiden09}
\begin{equation}
\sigma=\sigma_{0}+\sigma_{1}\left(\boldsymbol{P}^{\bar{p}}\cdot\boldsymbol{P}^{d}\right)+\left(\sigma_{2}-\sigma_{1}\right)\left(\boldsymbol{P}^{\bar{p}}\cdot\boldsymbol{v}\right)\left(\boldsymbol{P}^{d}\cdot\boldsymbol{v}\right)+\sigma_{3}P_{zz}^{d},\label{eq:PolarizedCrossSection}
\end{equation}
where $\boldsymbol{P}^{i}$~are the polarization vectors of corresponding
particles, $P_{zz}^{d}$~is the component of the deuteron tensor
polarization and $\boldsymbol{v}$~is the unit momentum vector. The
cross section $\sigma_{3}$ vanishes in the single-scattering approximation.
This cross section turned out to be much smaller than the cross sections
$\sigma_{1}$ and $\sigma_{2}$. This statement is valid also if the
shadowing effects are taken into account. The cross section $\sigma_{3}$
has no influence on the antiproton polarization and we neglect this
cross section in our further calculations. Spin-dependent parts of
the cross section can be expressed in terms of the scattering amplitudes~$g_{i}$
in the following way~\cite{UzikHaiden09}
\begin{equation}
\begin{array}{ll}
{\displaystyle \sigma_{0}=\vphantom{\Biggl|}\frac{2\pi}{k_{\bar{p}d}}\mathrm{Im}\left(g_{1}+g_{2}\right),\qquad\qquad\qquad} & {\displaystyle \sigma_{1}=\frac{4\pi}{k_{\bar{p}d}}\mathrm{Im}\, g_{3}},\\
{\displaystyle \sigma_{2}=\vphantom{\Biggl|}\frac{4\pi}{k_{\bar{p}d}}\mathrm{Im}\, g_{4}}, & {\displaystyle \sigma_{3}=\frac{4\pi}{k_{\bar{p}d}}\mathrm{Im}\frac{g_{1}-g_{2}}{6}}.
\end{array}
\end{equation}
Here
\begin{equation}
\begin{array}{l}
\vphantom{\biggl(}g_{1}=\hphantom{+}\frac{1}{2}\left(\left\langle +\frac{1}{2},-1\right|F^{\bar{p}d}\left(0\right)\left|+\frac{1}{2},-1\right\rangle +\left\langle +\frac{1}{2},+1\right|F^{\bar{p}d}\left(0\right)\left|+\frac{1}{2},+1\right\rangle \right),\\
\vphantom{\biggl(}g_{2}=\hphantom{-\frac{1}{\sqrt{2}}}\left\langle +\frac{1}{2},\hphantom{+}0\right|F^{\bar{p}d}\left(0\right)\left|+\frac{1}{2},\hphantom{+}0\right\rangle ,\\
\vphantom{\biggl(}g_{3}=-\frac{1}{\sqrt{2}}\left\langle +\frac{1}{2},-1\right|F^{\bar{p}d}\left(0\right)\left|-\frac{1}{2},\hphantom{+}0\right\rangle ,\\
\vphantom{\biggl(}g_{4}=\hphantom{+}\frac{1}{2}\left(\left\langle +\frac{1}{2},-1\right|F^{\bar{p}d}\left(0\right)\left|+\frac{1}{2},-1\right\rangle -\left\langle +\frac{1}{2},+1\right|F^{\bar{p}d}\left(0\right)\left|+\frac{1}{2},+1\right\rangle \right).
\end{array}
\end{equation}
In order to calculate these amplitudes we have substituted Eq.~\eqref{eq:ElasticAmplitude}
in the latter equations. One can see that it is necessary to calculate
the matrix elements of antiproton-nucleon scattering operators between
deuteron states with definite spin projections. A convenient way to
perform such calculations is to express the deuteron spin wave functions
via proton and neutron spin wave functions.

One can find the discussion of antiproton beam polarization buildup
in Refs.~\cite{MilStr05,DmitMilStr08}. Shown here is only the final
result for the polarization degree at time $t_{0}=2\tau_{b}$, $\tau_{b}=1/nf\sigma_{0}$
being the beam lifetime subject to scattering by the target:
\begin{equation}
P_{B}(t_{0})=\begin{cases}
-2P_{T}{\displaystyle \frac{\sigma_{1}}{\sigma_{0}}}, & \text{if}\:\boldsymbol{\zeta}_{T}\cdot\boldsymbol{v}=0,\\
-2P_{T}{\displaystyle \frac{\sigma_{2}}{\sigma_{0}}}, & \text{if}\,\left|\boldsymbol{\zeta}_{T}\cdot\boldsymbol{v}\right|=1.
\end{cases}\label{eq:BeamPolarization}
\end{equation}
Here $\boldsymbol{v}$~is the unit vector collinear to the antiproton
momentum, $\boldsymbol{\zeta}_{T}$~is the direction of the target
polarization, $P_{T}$~is the value of the target polarization, $n$~is
the areal density of the target and $f$~is the beam revolving frequency.
The equalities~\eqref{eq:BeamPolarization} are valid in both cases
$\boldsymbol{\zeta}_{T}\perp\boldsymbol{v}$ and $\boldsymbol{\zeta}_{T}\parallel\boldsymbol{v}$.

\section{Results\label{sec:Results}}

\begin{figure}[b]
\begin{centering}
\hspace{7mm}\includegraphics[height=4.5cm]{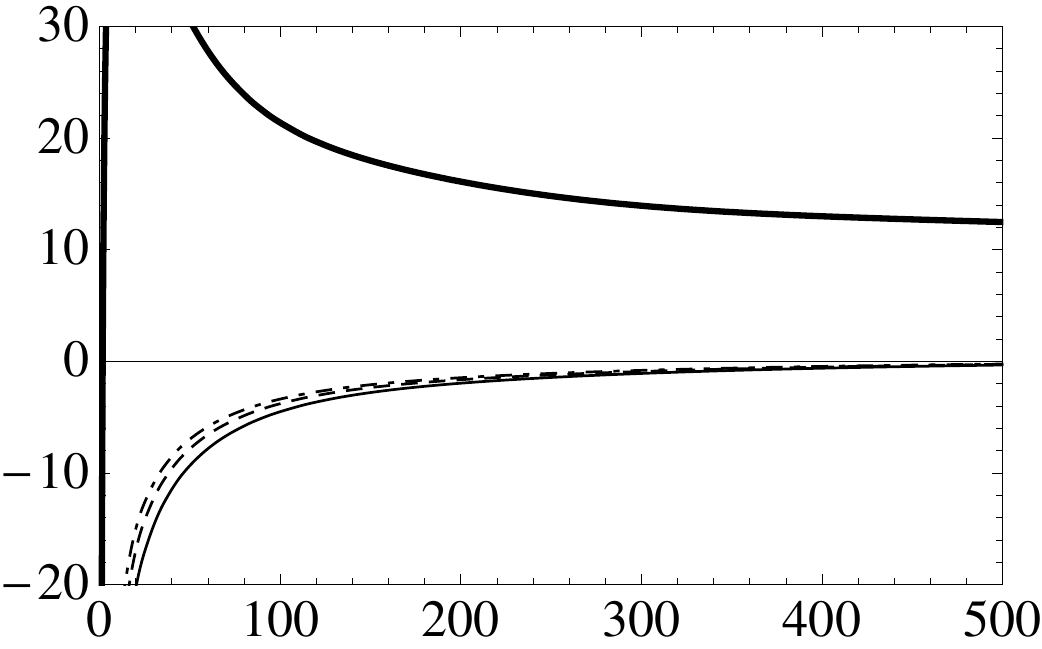}\hfill{}\includegraphics[height=4.5cm]{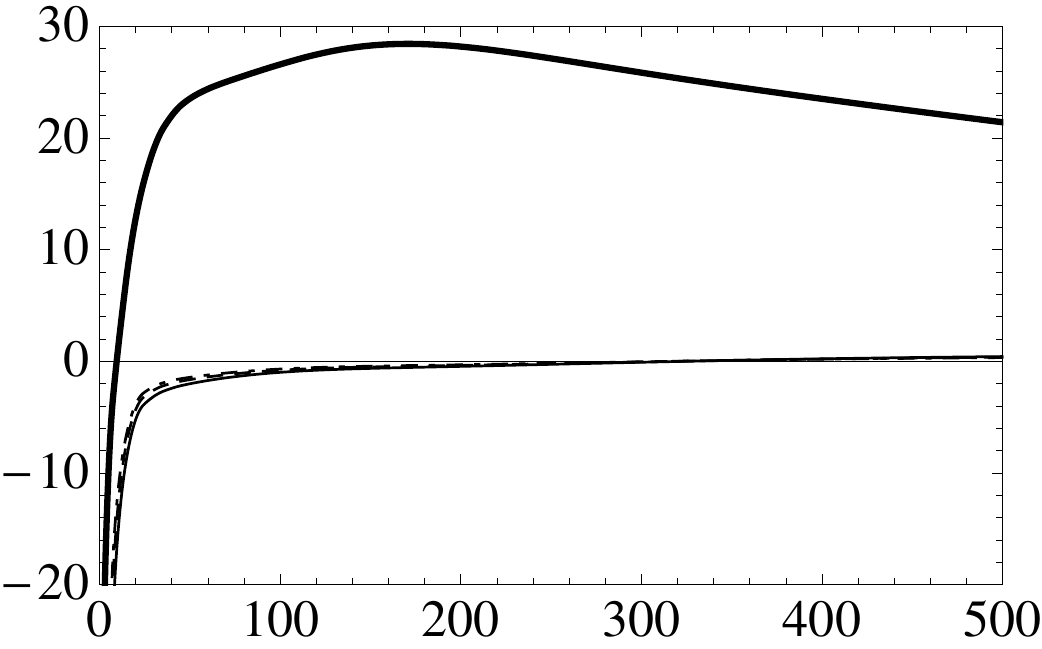}
\par\end{centering}

\begin{centering}
\begin{picture}(0,0)(0,0) \put(-135,10){ $T_{\mathrm{lab}}\,(\mathrm{MeV})$}
\put(110,10){ $T_{\mathrm{lab}}\,(\mathrm{MeV})$} \put(-230,85){\rotatebox{90}{$\sigma_{1}\,(\mathrm{mb})$}}
\put(10,85){\rotatebox{90}{$\sigma_{2}\,(\mathrm{mb})$}} \put(-42,90){
{\large $\bar{p}p$}} \put(200,90){ {\large $\bar{p}p$}} \end{picture}
\par\end{centering}

\begin{centering}
\hspace{10mm}\includegraphics[height=4.5cm]{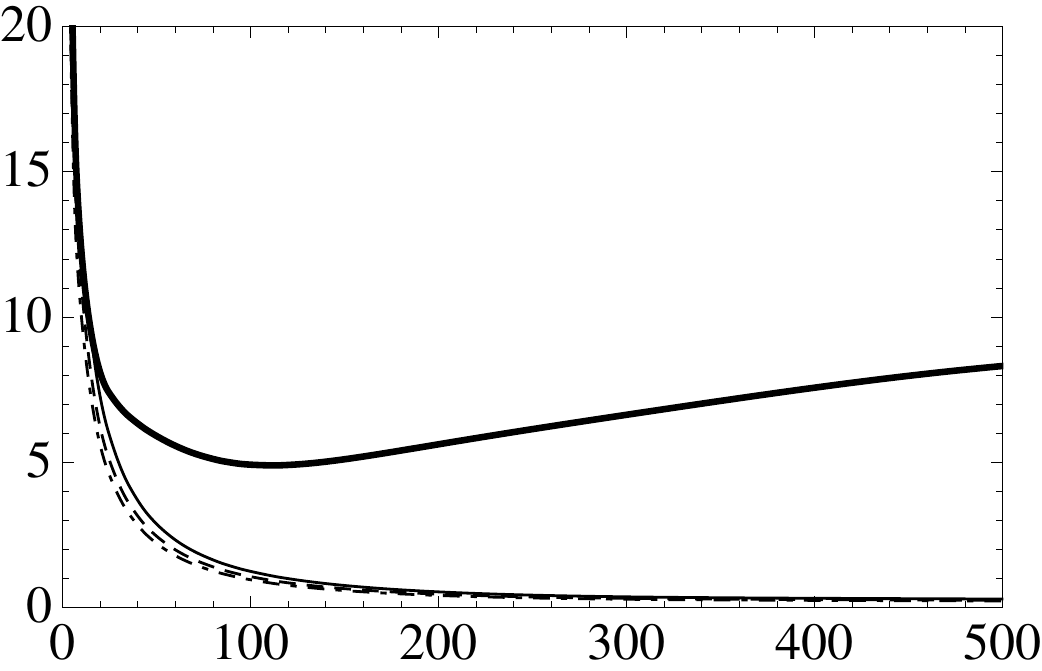}\hfill{}\includegraphics[height=4.5cm]{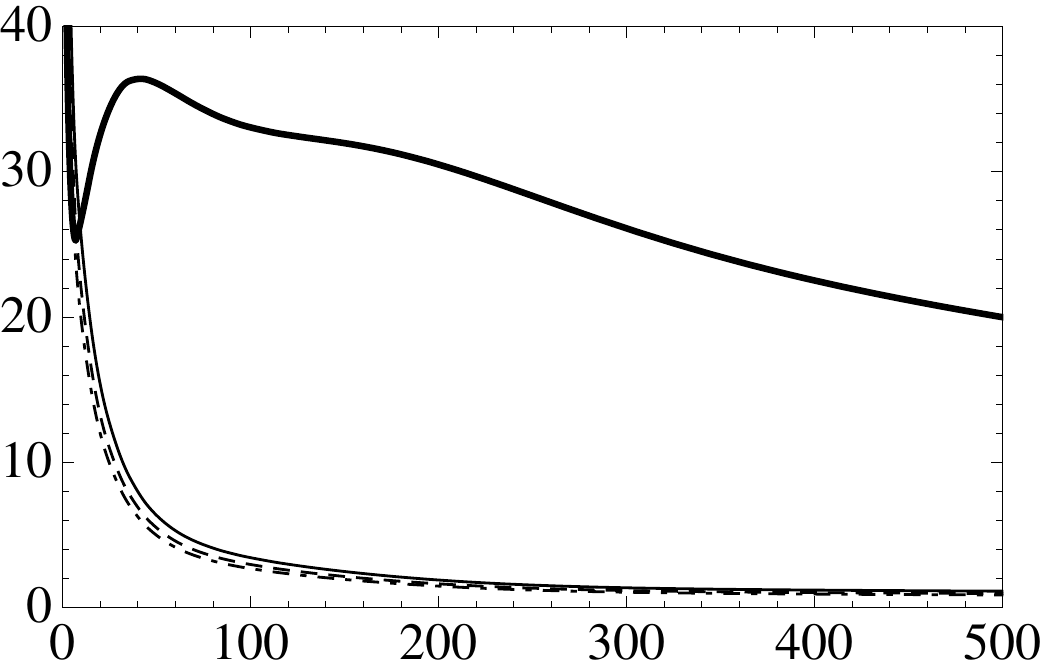}
\par\end{centering}

\begin{centering}
\begin{picture}(0,0)(0,0) \put(-135,10){ $T_{\mathrm{lab}}\,(\mathrm{MeV})$}
\put(110,10){ $T_{\mathrm{lab}}\,(\mathrm{MeV})$} \put(-230,85){\rotatebox{90}{$\sigma_{1}\,(\mathrm{mb})$}}
\put(10,85){\rotatebox{90}{$\sigma_{2}\,(\mathrm{mb})$}} \put(-42,50){
{\large $\bar{p}n$}} \put(200,50){ {\large $\bar{p}n$}} \end{picture}
\par\end{centering}

\caption{\label{fig:Sigma_p_n}The dependence of hadronic cross sections $\sigma_{1}^{\mathrm{h}}$,~$\sigma_{2}^{\mathrm{h}}$
(thick line) and the Coulomb-hadronic interference contributions $\sigma_{1}^{\mathrm{int}}$,~$\sigma_{2}^{\mathrm{int}}$
(thin lines) on $T_{\mathrm{lab}}$ for $\bar{p}p$ scattering (upper
row) and $\bar{p}n$ scattering (lower row). The interference contribution
in the lower row is the interference of the Coulomb $\bar{p}p$ and
strong $\bar{p}n$ amplitudes for $\bar{p}d$ scattering. The acceptance
angles in the lab frame are \mbox{$\theta_{\mathrm{acc}}^{(l)}=\unit[10]{mrad}$}
(solid line), \mbox{$\theta_{\mathrm{acc}}^{(l)}=\unit[20]{mrad}$}
(dashed line), \mbox{$\theta_{\mathrm{acc}}^{(l)}=\unit[30]{mrad}$}
(dashed\protect\nobreakdash-dotted line).}
\end{figure}
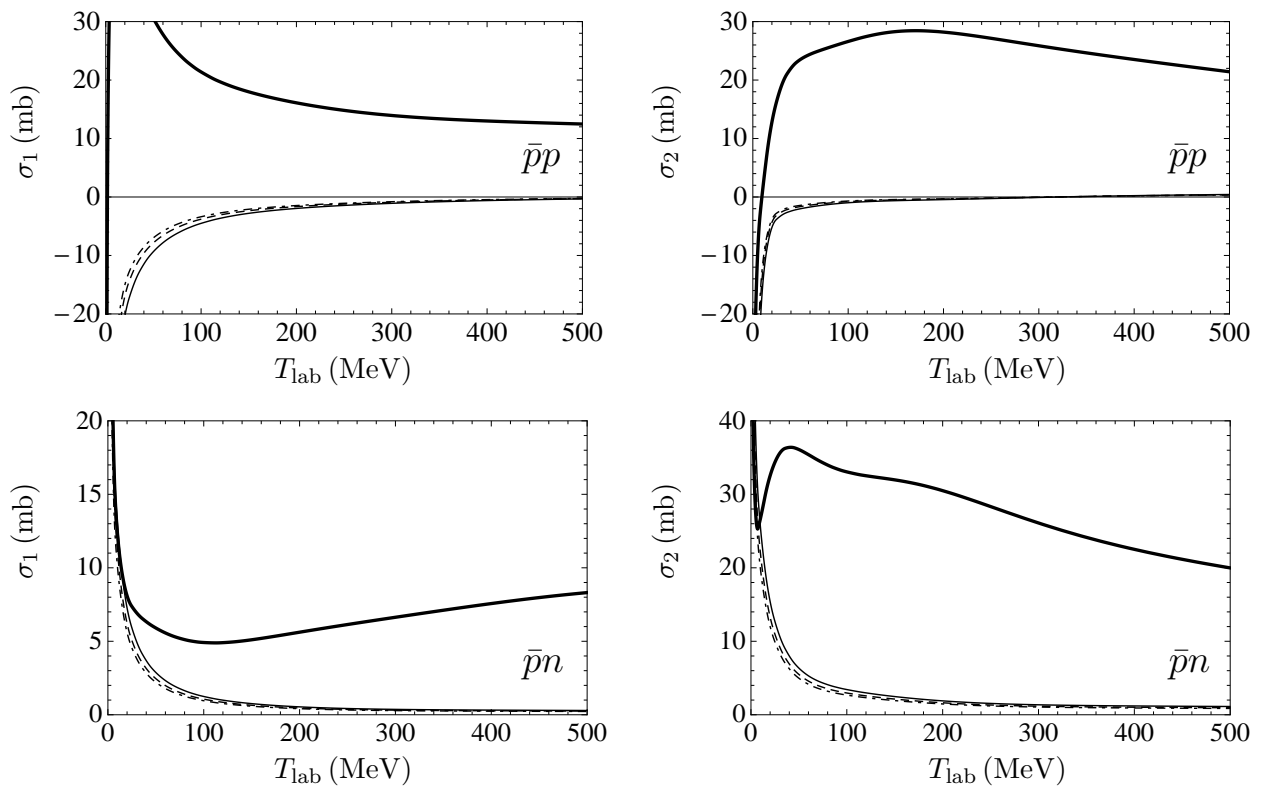

In this section we present our numerical results for the spin-dependent
parts of the cross sections of $\bar{p}p$, $\bar{p}n$ and $\bar{p}d$
scattering along with the predictions for the beam polarization degree.
Our results for total unpolarized $\bar{p}p$ and $\bar{p}n$ cross
sections are in good agreement with all available experimental data.
Unpolarized cross sections were studied both theoretically and experimentally
by many authors so there is no necessity to present the corresponding
Figures here. However, the situation is different for the spin-dependent
parts of the cross sections because they were not studied experimentally
and different theoretical models provide essentially different predictions.
Our predictions for the spin-dependent parts of $\bar{p}p$ cross
section were previously presented in Ref.~\cite{DmitMilSal10}, but
we show them here for completeness. The dependence of the spin-dependent
parts of the cross section of $\bar{p}p$ and $\bar{p}n$ scattering
is shown in Fig.~\ref{fig:Sigma_p_n}. One can see that $\sigma_{2}$
is of the same order for these two processes, but $\sigma_{1}$ for
$\bar{p}n$ scattering is smaller than that for $\bar{p}p$ scattering.
Note that the sing of the interference contribution to $\bar{p}n$
scattering differs from that to $\bar{p}p$ scattering.

\begin{figure}
\hspace{4.4cm}(a)\hfill{}(b)\hspace{3.2cm}

\begin{centering}
\hspace{9mm}\includegraphics[height=4.5cm]{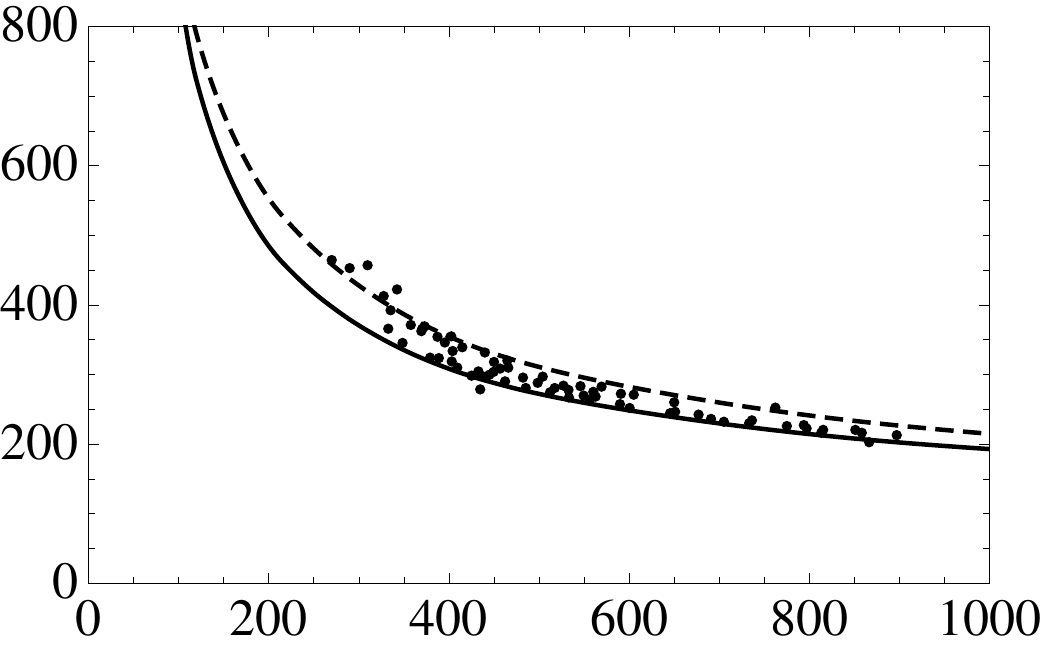}\hfill{}\includegraphics[height=4.5cm]{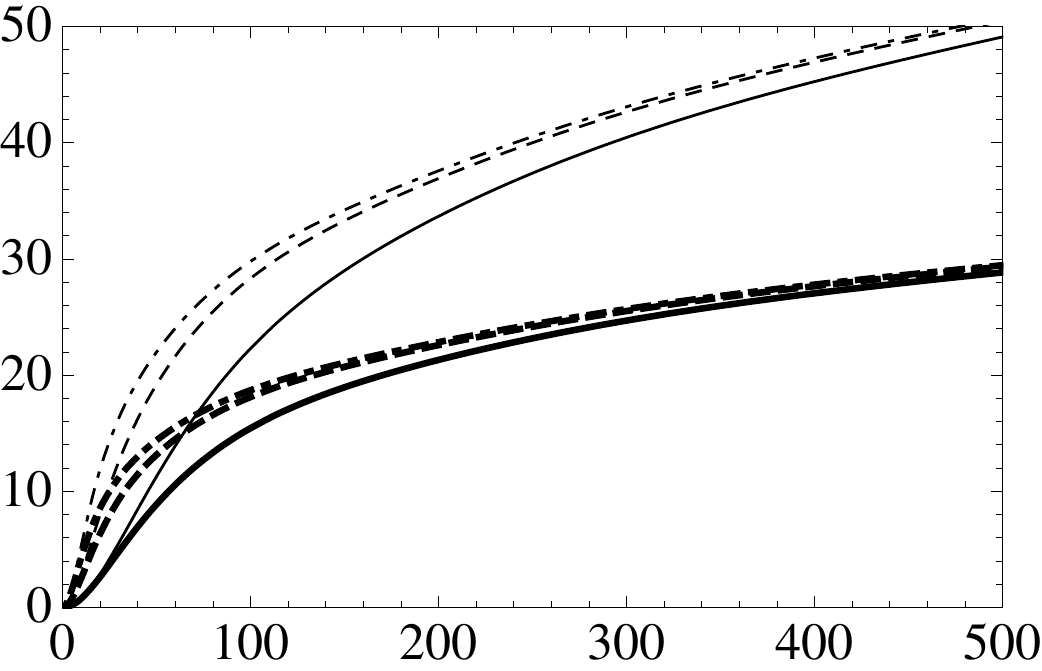}
\par\end{centering}

\begin{centering}
\begin{picture}(0,0)(0,0) \put(-135,10){ $p\,(\mathrm{MeV}/c)$}
\put(-230,85){\rotatebox{90}{$\sigma_{\mathrm{tot}}\,(\mathrm{mb})$}}
\put(110,10){ $T_{\mathrm{lab}}\,(\mathrm{MeV})$} \put(10,85){\rotatebox{90}{$t_{0}\,(\mathrm{hour})$}}
\end{picture}
\par\end{centering}

\caption{\label{fig:Sigma_pd_total_t0}(a) The dependence of the total $\bar{p}d$
cross section on $p_{\mathrm{lab}}$ within the single-scattering
approximation (dashed line) and including shadowing effects (solid
line). Data are taken from Refs.~\cite{Bizzarri74,KalogTzan80,Burrows70,Carroll74,HamPunTrippLazNich80}.
(b) The dependence of the polarization time $t_{0}$ on $T_{\mathrm{lab}}$
for $n=\unit[10^{14}]{cm^{-2}}$ and $f=\unit[10^{6}]{c^{-1}}$ for
$\bar{p}d$ scattering (thick lines) and $\bar{p}p$ scattering (thin
lines). The acceptance angles in the lab frame are \mbox{$\theta_{\mathrm{acc}}^{(l)}=\unit[10]{mrad}$}
(solid line), \mbox{$\theta_{\mathrm{acc}}^{(l)}=\unit[20]{mrad}$}
(dashed line), \mbox{$\theta_{\mathrm{acc}}^{(l)}=\unit[30]{mrad}$}
(dashed\protect\nobreakdash-dotted line).}
\end{figure}
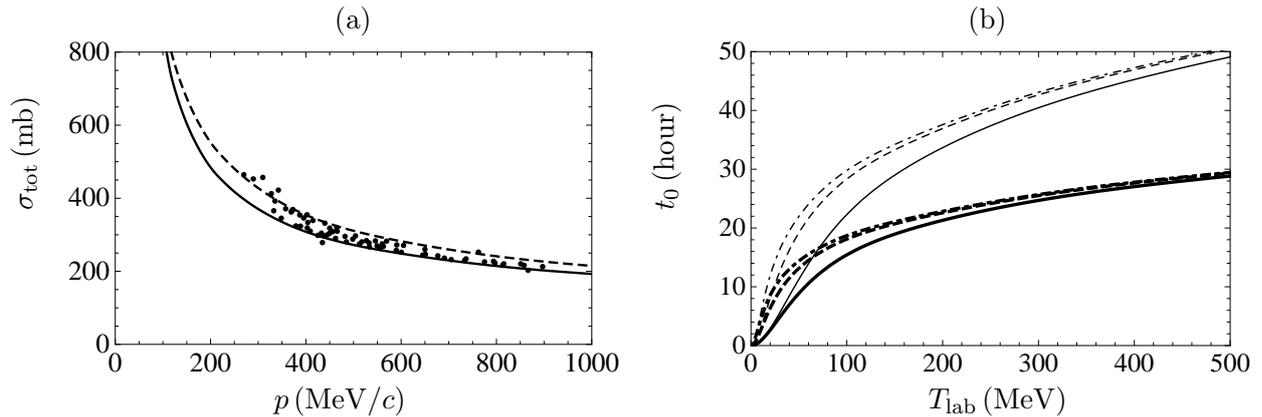

In order to estimate the role of double-scattering mechanism in $\bar{p}d$
scattering we have calculated the total unpolarized $\bar{p}d$ cross
section (see~Fig.~\ref{fig:Sigma_pd_total_t0}-a). One can see that
the shadowing effects decrease the total cross section at about $15\%$
in the whole energy region. The line obtained in this work approximates
the experimental data~\cite{Bizzarri74,KalogTzan80,Burrows70,Carroll74,HamPunTrippLazNich80}
quite accurately.

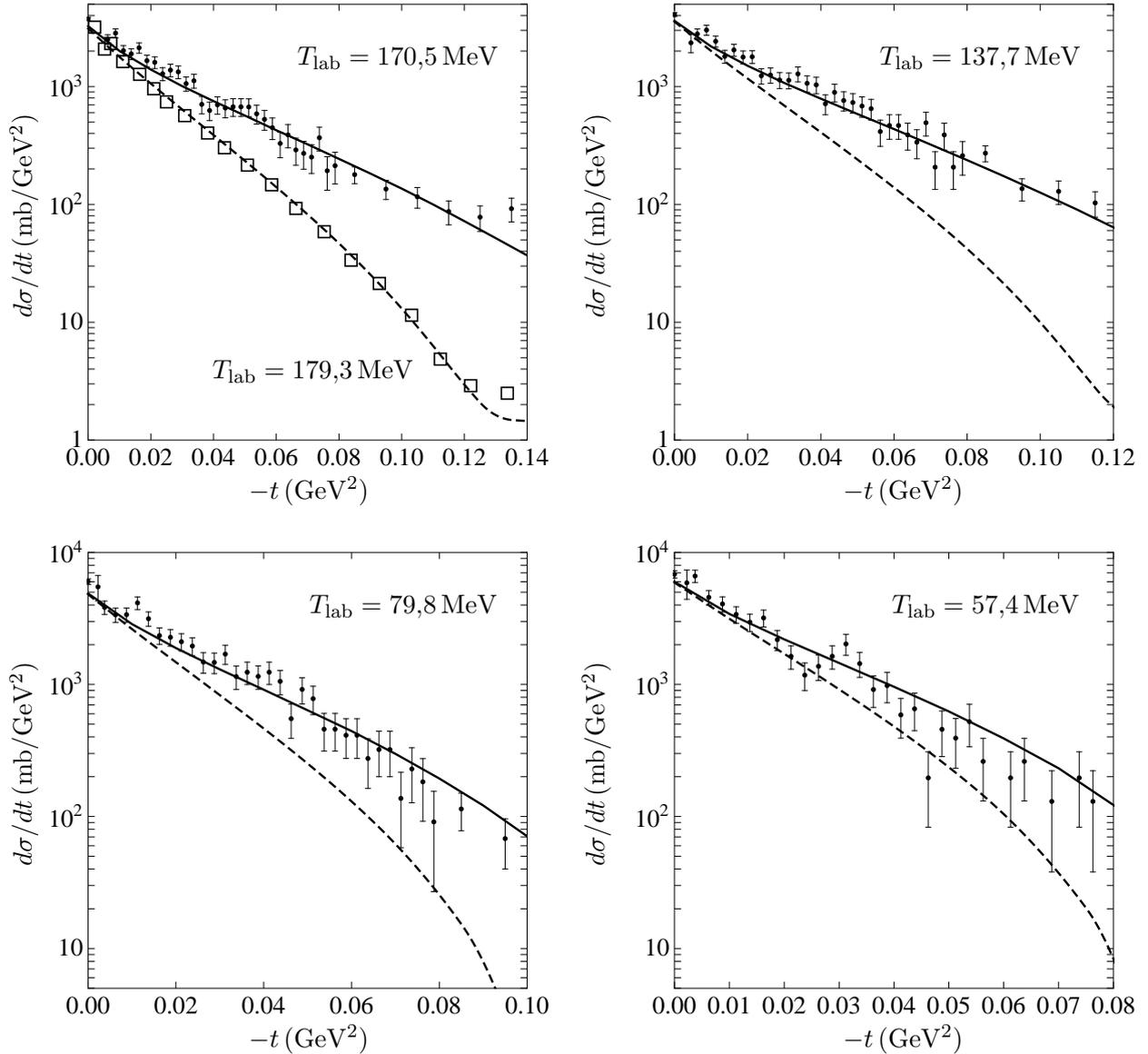
\begin{figure}
\begin{centering}
\hspace{7mm}\includegraphics[height=7.2cm]{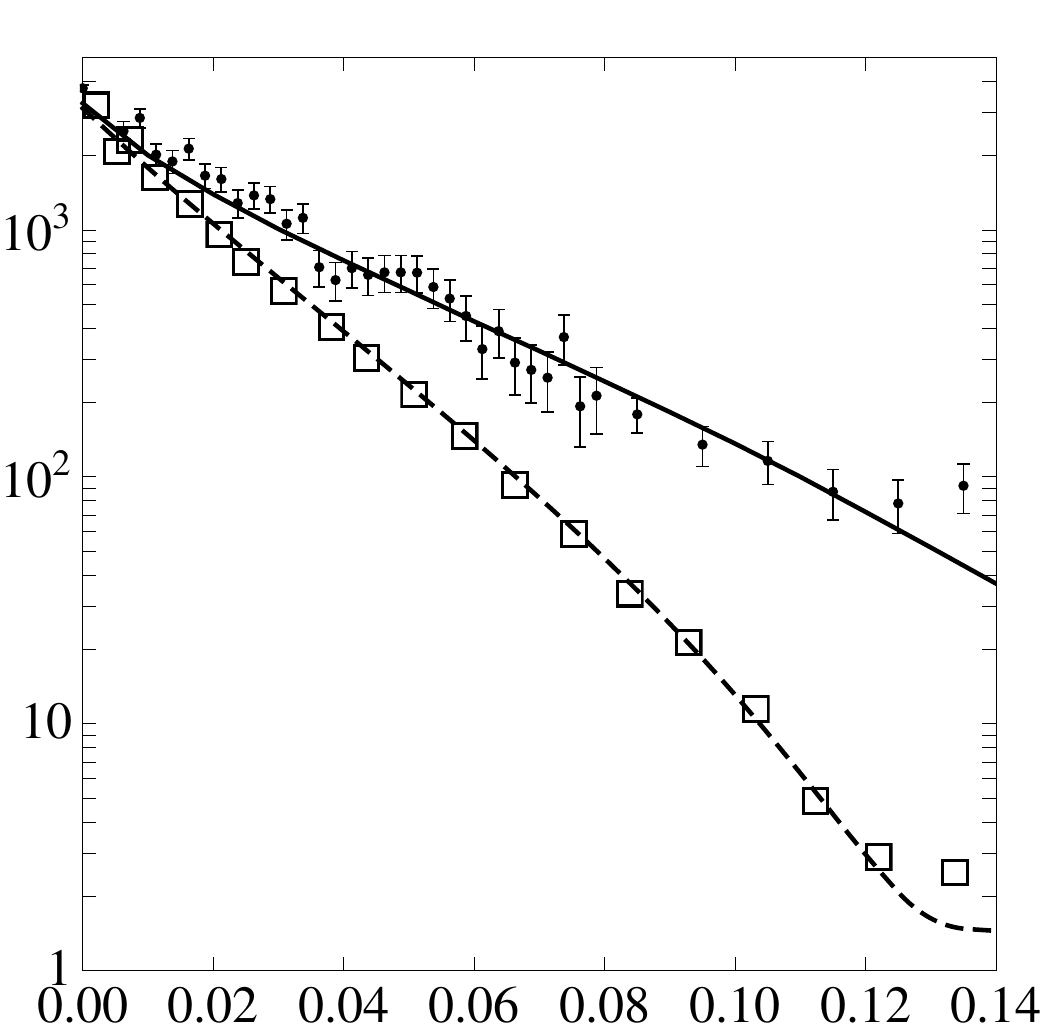}\hfill{}\includegraphics[height=7.2cm]{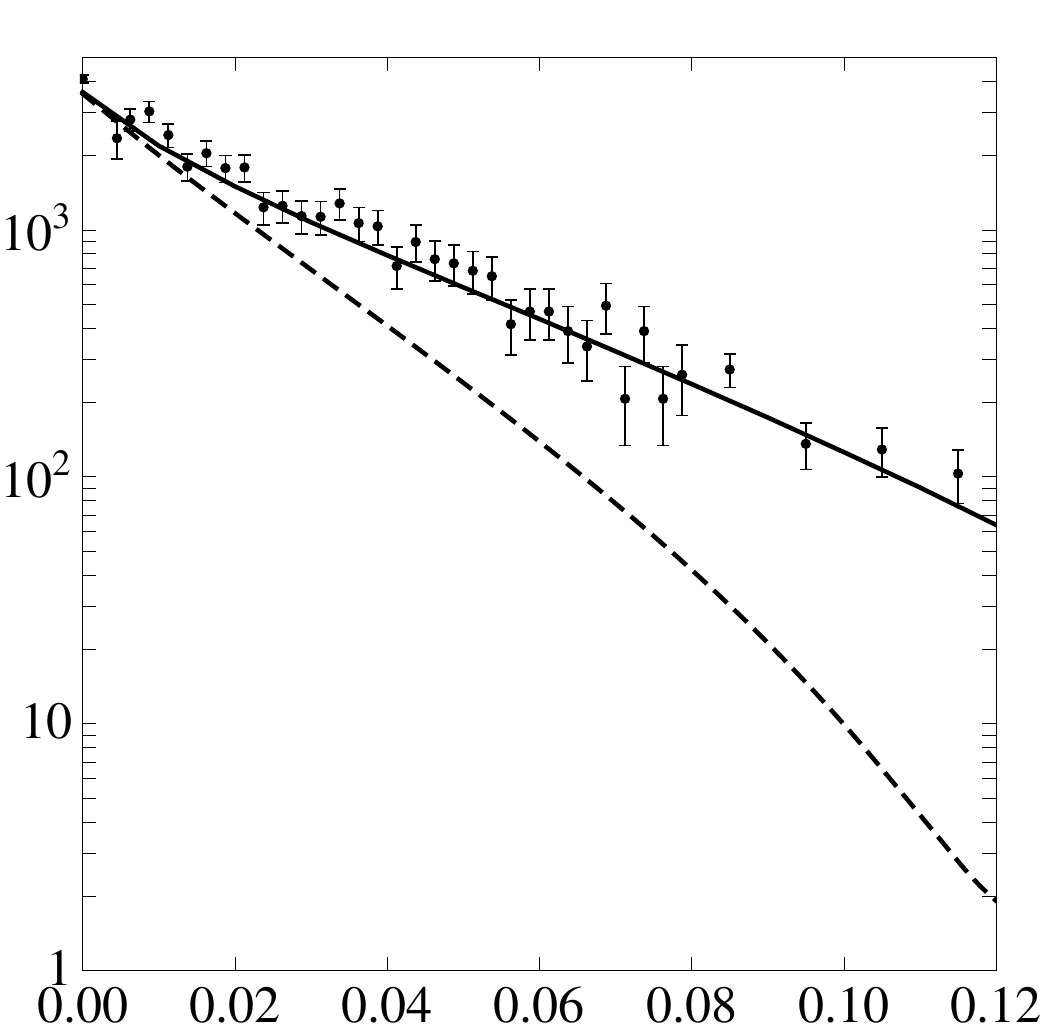}
\par\end{centering}

\begin{centering}
\begin{picture}(0,0)(0,0) \put(-135,10){ $-t\,(\mathrm{GeV^{2}})$}
\put(110,10){ $-t\,(\mathrm{GeV^{2}})$} \put(-230,85){\rotatebox{90}{$d\sigma/dt\,(\mathrm{mb/GeV^{2}})$}}
\put(6,85){\rotatebox{90}{$d\sigma/dt\,(\mathrm{mb/GeV^{2}})$}}
\put(-115,190){ $T_{\mathrm{lab}}=\unit[170{,}5]{MeV}$} \put(-150,60){
$T_{\mathrm{lab}}=\unit[179{,}3]{MeV}$} \put(125,190){ $T_{\mathrm{lab}}=\unit[137{,}7]{MeV}$}
\end{picture}
\par\end{centering}

\begin{centering}
\hspace{7mm}\includegraphics[height=7.2cm]{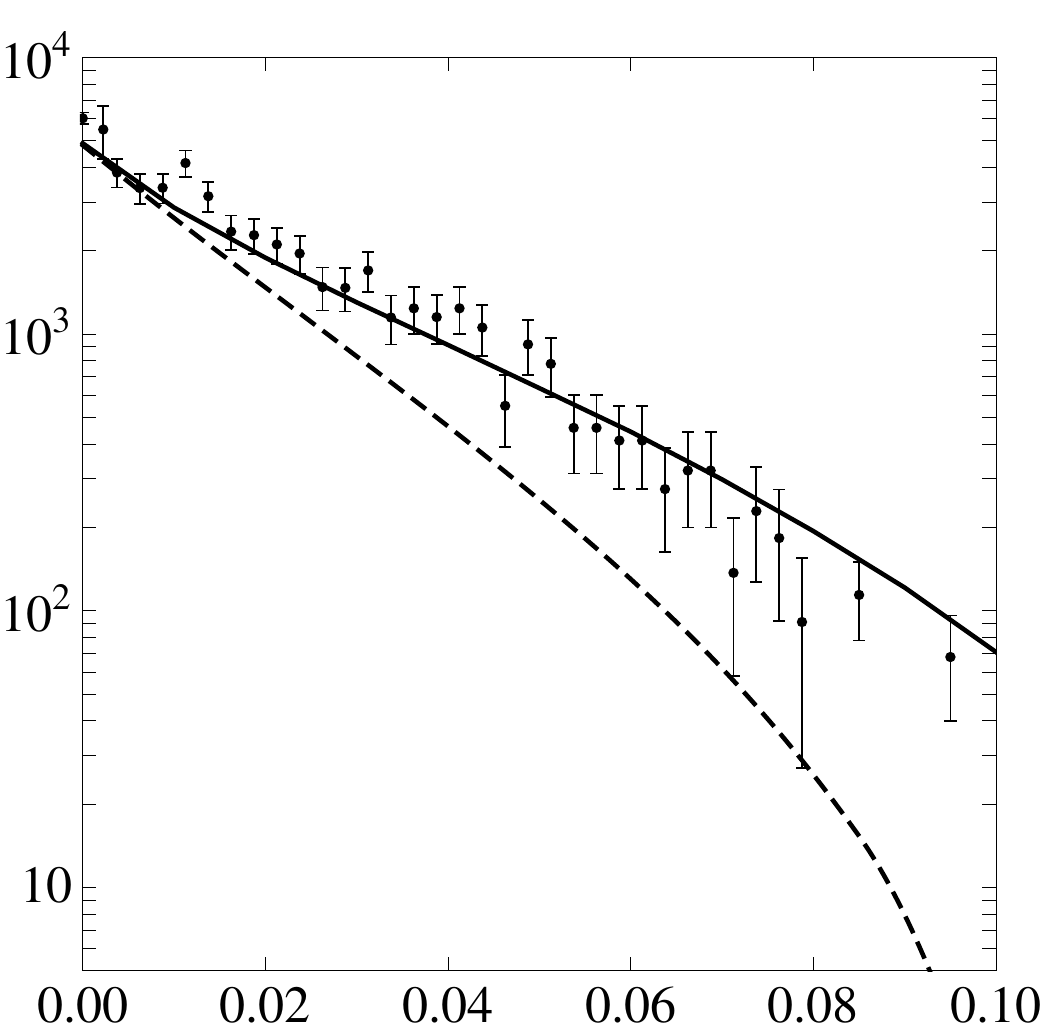}\hfill{}\includegraphics[height=7.2cm]{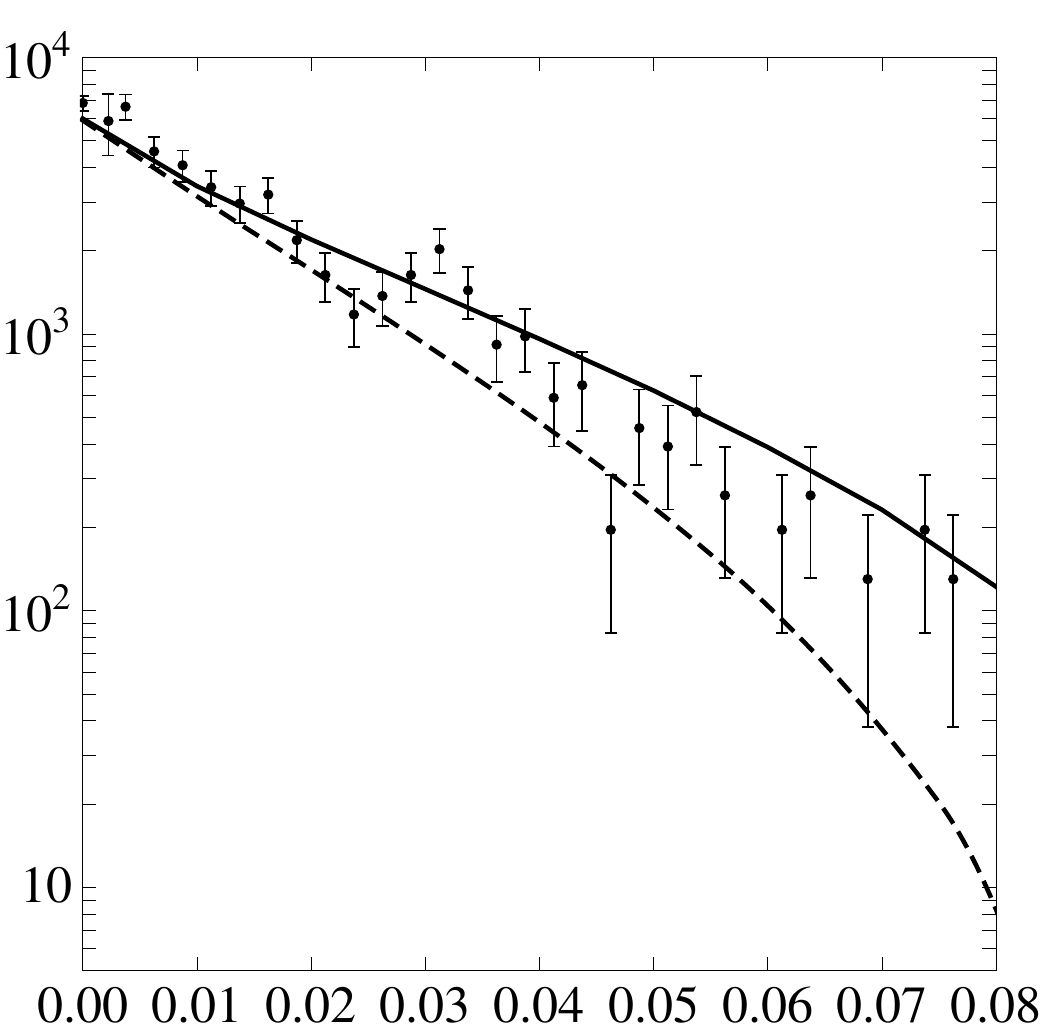}
\par\end{centering}

\begin{centering}
\begin{picture}(0,0)(0,0) \put(-135,10){ $-t\,(\mathrm{GeV^{2}})$}
\put(110,10){ $-t\,(\mathrm{GeV^{2}})$} \put(-230,85){\rotatebox{90}{$d\sigma/dt\,(\mathrm{mb/GeV^{2}})$}}
\put(6,85){\rotatebox{90}{$d\sigma/dt\,(\mathrm{mb/GeV^{2}})$}}
\put(-110,190){ $T_{\mathrm{lab}}=\unit[79{,}8]{MeV}$} \put(130,190){
$T_{\mathrm{lab}}=\unit[57{,}4]{MeV}$} \end{picture}
\par\end{centering}

\caption{\label{fig:Sigma_diff}The dependence of elastic (dashed lines) and
elastic plus inelastic (solid lines) $\bar{p}d$ differential cross
sections on the momentum transfer. Data for the elastic scattering
cross section (squares) are taken from Ref.~\cite{Bruge88} and for
elastic plus inelastic cross sections (dots) from Ref.~\cite{Bizzarri74}.}
\end{figure}

We also present here the differential elastic ($\bar{p}d\to\bar{p}d$)
and elastic plus inelastic ($\bar{p}d\to\bar{p}\, pn$)  cross sections
(see~Fig.~\ref{fig:Sigma_diff}). These quantities are interesting
for us because the double-scattering mechanism is very important for
accurate description of non-forward $\bar{p}d$ scattering and we
can test the applicability of the Glauber theory to low-energy $\bar{p}d$
scattering comparing our predictions with the existing experimental
data. We have included the D-wave contribution with the method described
in Ref.~\cite{FrancoGlauber69} while calculating the elastic differential
cross sections. In order to calculate the two form-factors needed
by the theory, the numerical values for the deuteron wave function
calculated in Ref.~\cite{Lacombe80} using the Paris model were used.
As we expected, the D-wave contribution proved to be significant only
for scattering with large momentum transfer because the corresponding
form-factor vanishes in the case of forward scattering. Experimental
data for the elastic $\bar{p}d$ scattering exist only at $\unit[180]{MeV}$~\cite{Bruge88}
(squares in Fig.~\ref{fig:Sigma_diff}) and are nicely reproduced
by our line. Experimental data for elastic plus inelastic scattering~\cite{Bizzarri74}
are also reproduced quite well, see Fig.~\ref{fig:Sigma_diff}. The
Glauber theory seems to be applicable for the description of unpolarized
$\bar{p}d$ cross sections at rather low energies down to $\unit[50]{MeV}$.

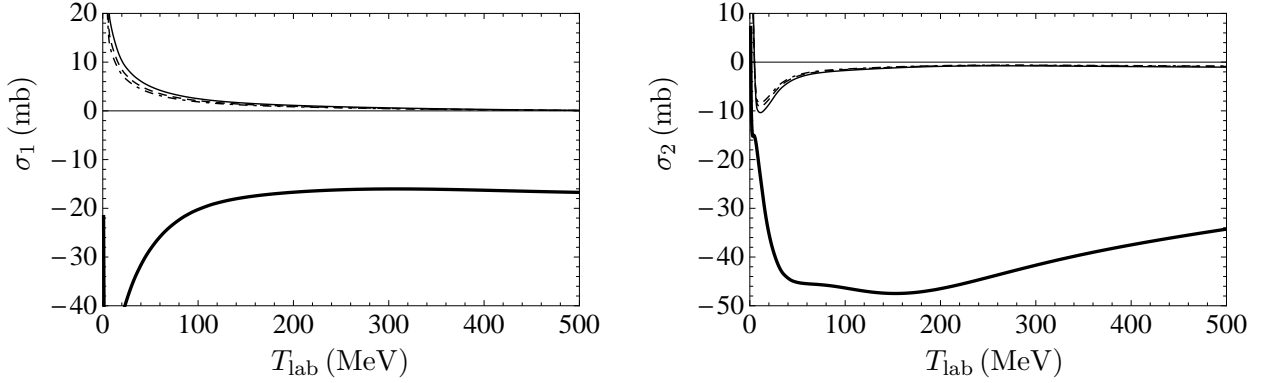
\begin{figure}
\begin{centering}
\hspace{7mm}\includegraphics[height=4.5cm]{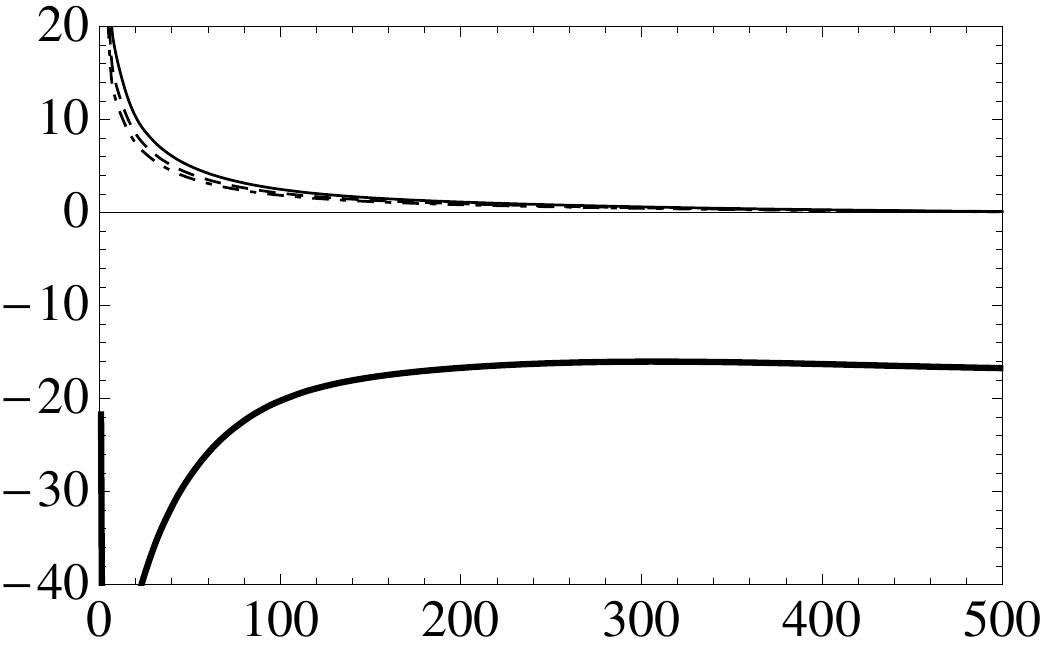}\hfill{}\includegraphics[height=4.5cm]{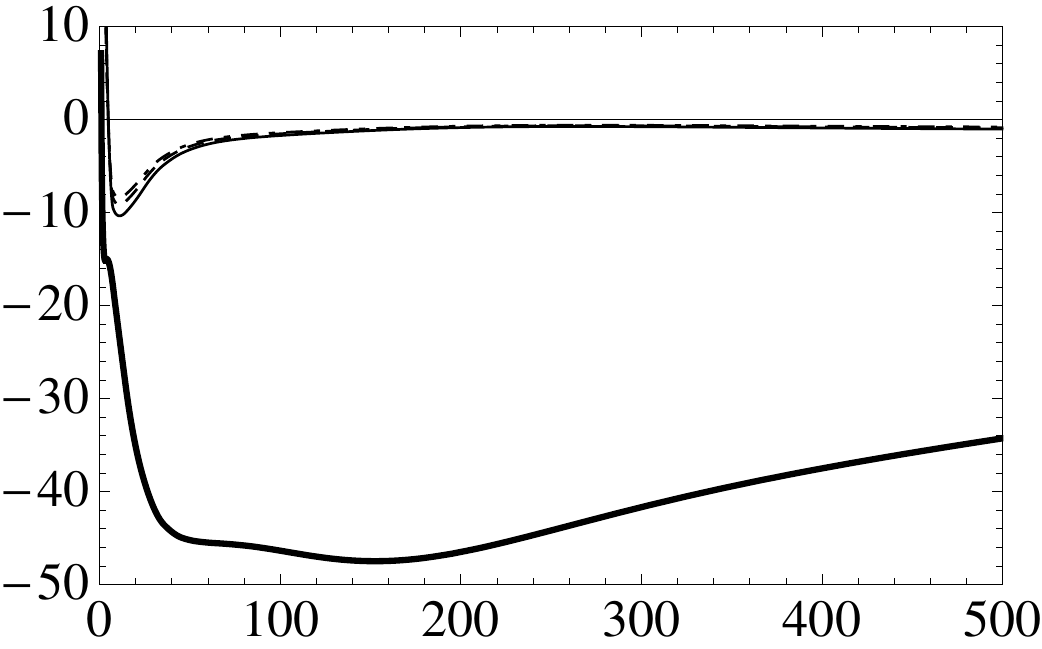}
\par\end{centering}

\begin{centering}
\begin{picture}(0,0)(0,0) \put(-135,10){ $T_{\mathrm{lab}}\,(\mathrm{MeV})$}
\put(110,10){ $T_{\mathrm{lab}}\,(\mathrm{MeV})$} \put(-230,85){\rotatebox{90}{$\sigma_{1}\,(\mathrm{mb})$}}
\put(10,85){\rotatebox{90}{$\sigma_{2}\,(\mathrm{mb})$}} \end{picture}
\par\end{centering}

\caption{\label{fig:Sigma_d}The dependence of $\sigma_{1}$, $\sigma_{2}$
(thick line) and interference contributions $\sigma_{1}^{\mathrm{int}}$,~$\sigma_{2}^{\mathrm{int}}$
(thin lines) on $T_{\mathrm{lab}}$ for $\bar{p}d$ scattering. The
acceptance angles in the lab frame are \mbox{$\theta_{\mathrm{acc}}^{(l)}=\unit[10]{mrad}$}
(solid line), \mbox{$\theta_{\mathrm{acc}}^{(l)}=\unit[20]{mrad}$}
(dashed line), \mbox{$\theta_{\mathrm{acc}}^{(l)}=\unit[30]{mrad}$}
(dashed\protect\nobreakdash-dotted line).}
\end{figure}

The spin-dependent parts of the cross section of $\bar{p}d$ scattering
were calculated with the double-scattering mechanism taken into account.
However, the contribution of D-wave in deuteron wave function was
omitted because we expect it to be less important. The spin-dependent
parts of $\bar{p}d$ cross section are presented in Fig.~\ref{fig:Sigma_d}.
The interference contribution to $\bar{p}d$ cross section proved
to be less significant in most part of the energy range than it was
for $\bar{p}p$ scattering. Shadowing effects turned out to decrease
the absolute value of cross sections $\sigma_{1}$ and $\sigma_{2}$
at about $20\div25\%$ level.

\begin{figure}
\begin{centering}
\hspace{7mm}\includegraphics[height=4.5cm]{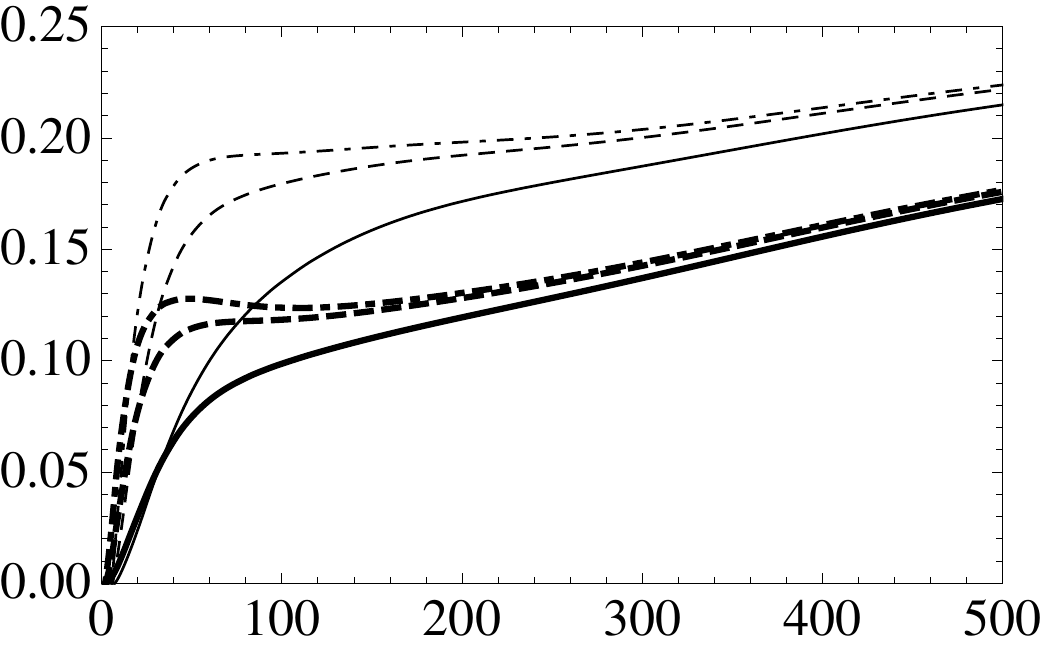}\hfill{}\includegraphics[height=4.35cm]{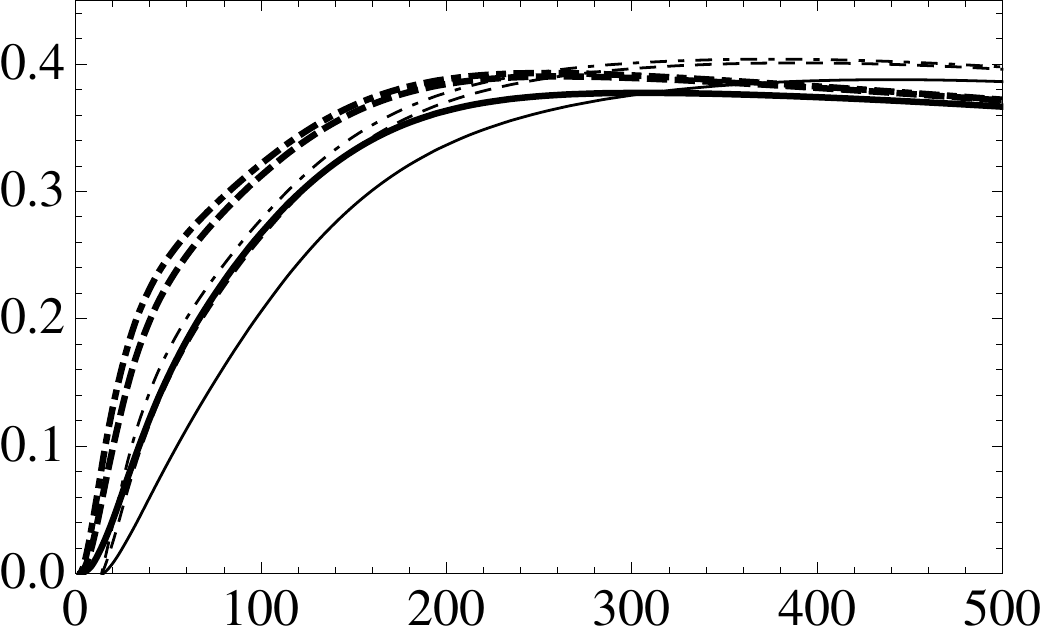}
\par\end{centering}

\begin{centering}
\begin{picture}(0,0)(0,0) \put(-135,10){ $T_{\mathrm{lab}}\,(\mathrm{MeV})$}
\put(110,10){ $T_{\mathrm{lab}}\,(\mathrm{MeV})$} \put(-230,85){\rotatebox{90}{$P_{\perp}$}}
\put(10,85){\rotatebox{90}{$P_{\parallel}$}} \end{picture}
\par\end{centering}

\begin{centering}
\hspace{7mm}\includegraphics[height=4.5cm]{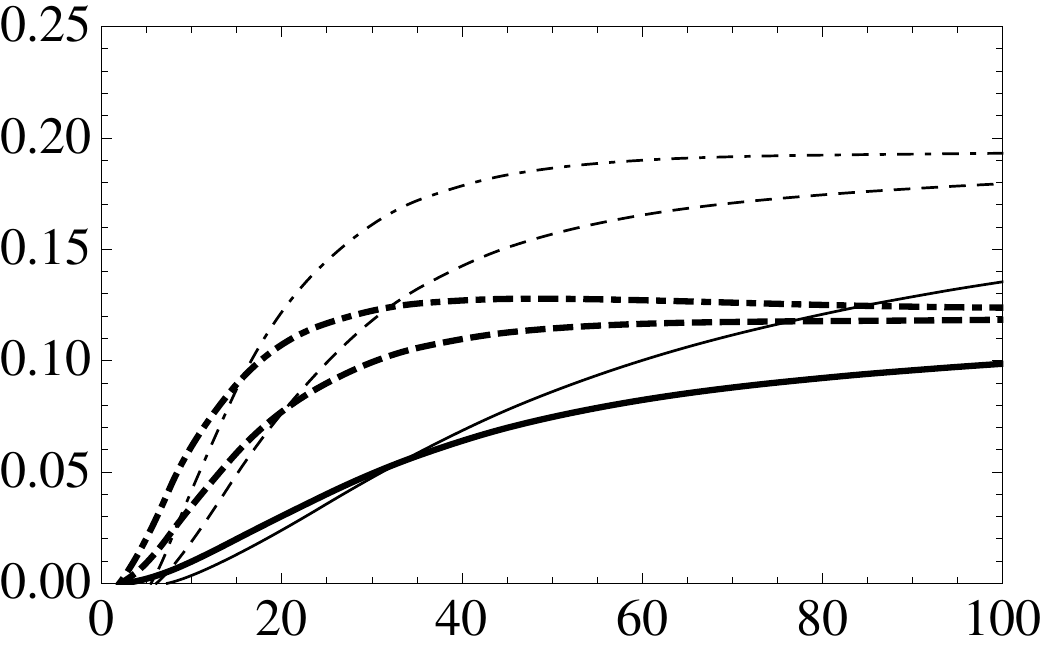}\hfill{}\includegraphics[height=4.35cm]{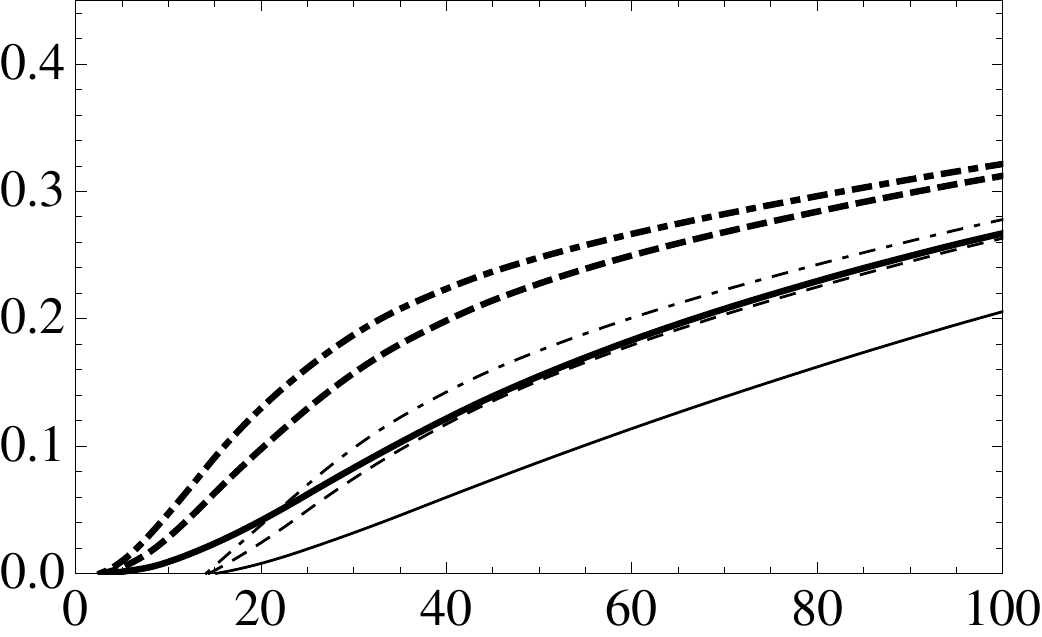}
\par\end{centering}

\begin{centering}
\begin{picture}(0,0)(0,0) \put(-135,10){ $T_{\mathrm{lab}}\,(\mathrm{MeV})$}
\put(110,10){ $T_{\mathrm{lab}}\,(\mathrm{MeV})$} \put(-230,85){\rotatebox{90}{$P_{\perp}$}}
\put(10,85){\rotatebox{90}{$P_{\parallel}$}} \end{picture}
\par\end{centering}

\caption{\label{fig:Polarization_d}The dependence of $P_{B}(t_{0})$ for $P_{T}=1$
on $T_{\mathrm{lab}}$ for \mbox{$\boldsymbol{\zeta}_{T}\cdot\boldsymbol{v}=0$
($P_{\perp}$)} and \mbox{$\left|\boldsymbol{\zeta}_{T}\cdot\boldsymbol{v}\right|=1$}
($P_{\|}$) for $\bar{p}d$ scattering (thick lines) in comparison
with $\bar{p}p$ scattering (thin lines). Note that the polarization
degree for $\bar{p}p$ scattering is shown with the opposite sign
for simplicity. The acceptance angles in the lab frame are \mbox{$\theta_{\mathrm{acc}}^{(l)}=\unit[10]{mrad}$}
(solid lines), \mbox{$\theta_{\mathrm{acc}}^{(l)}=\unit[20]{mrad}$}
(dashed lines), \mbox{$\theta_{\mathrm{acc}}^{(l)}=\unit[30]{mrad}$}
(dashed\protect\nobreakdash-dotted lines). Low energy region is shown
again in the lower row.}
\end{figure}

Let us proceed now to the discussion of the polarization buildup.
The dependence of the time of polarization $t_{0}$ on the antiproton
energy is presented in Fig.~\ref{fig:Sigma_pd_total_t0}-b. Note
that the number of antiprotons at time $t_{0}$ equals to $14\%$
of the initial number. The dependence of transverse and longitudinal
polarization degrees on the antiproton energy is shown in Fig.~\ref{fig:Polarization_d}.
Analogous results from Ref.~\cite{DmitMilSal10} for $\bar{p}p$
scattering are also shown in that Figure with the thin lines. One
can see that the transverse polarization in the case of deuterium
target is smaller than that in the case of hydrogen target. However,
it is almost the same for energies below $\unit[50]{MeV}$. The picture
is different for longitudinal polarization. The longitudinal polarization
degree in the case of $\bar{p}d$ scattering is larger for low energies,
but it is almost the same as for $\bar{p}p$ scattering in most of
energy range concerned.

\begin{figure}
\begin{centering}
\hspace{5mm}\includegraphics[height=4.6cm]{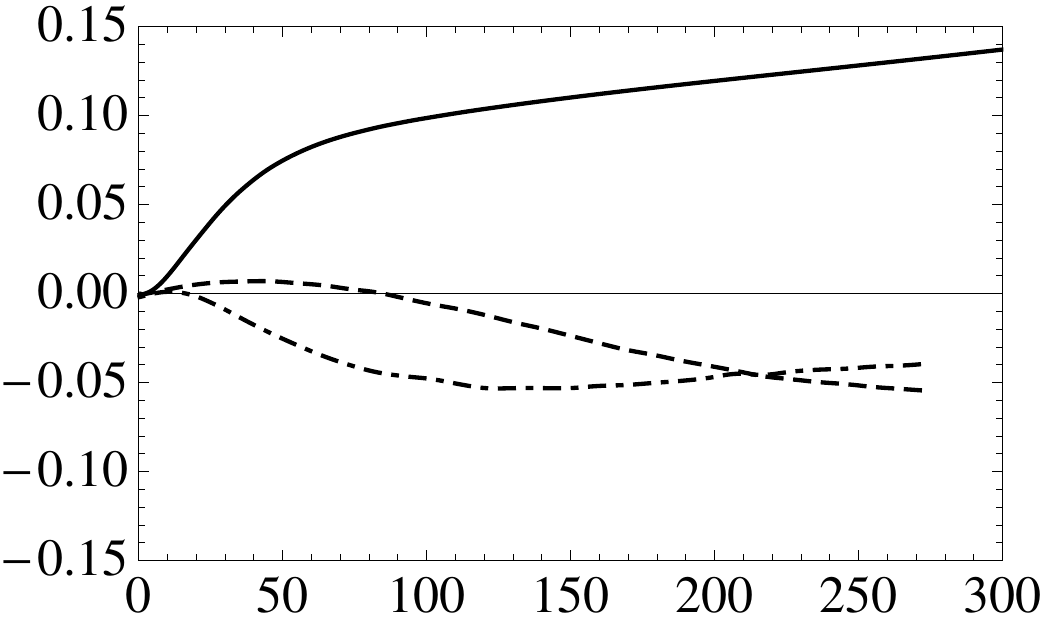}\hfill{}\includegraphics[height=4.6cm]{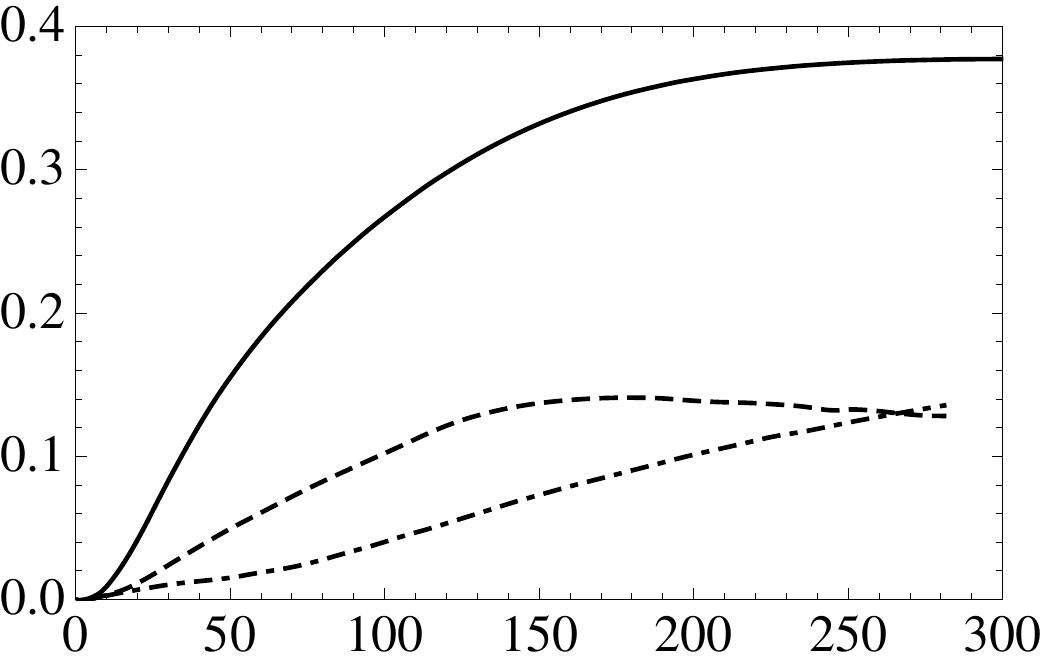}
\par\end{centering}

\begin{centering}
\begin{picture}(0,0)(0,0) \put(-135,10){ $T_{\mathrm{lab}}\,(\mathrm{MeV})$}
\put(110,10){ $T_{\mathrm{lab}}\,(\mathrm{MeV})$} \put(-232,95){\rotatebox{90}{$P_{\perp}$}}
\put(10,95){\rotatebox{90}{$P_{\parallel}$}} \end{picture}
\par\end{centering}

\caption{\label{fig:Polarization_d_compare}The dependence of $P_{B}(t_{0})$
for $P_{T}=1$ on $T_{\mathrm{lab}}$ for \mbox{$\boldsymbol{\zeta}_{T}\cdot\boldsymbol{v}=0$
($P_{\perp}$)} and \mbox{$\left|\boldsymbol{\zeta}_{T}\cdot\boldsymbol{v}\right|=1$}
($P_{\|}$) for $\bar{p}d$ scattering for different models: Nijmegen
model (solid line), Julich A model (dashed line) and Julich D model
(dashed\protect\nobreakdash-dotted line). The acceptance angle in
the lab frame is \mbox{$\theta_{\mathrm{acc}}^{(l)}=\unit[10]{mrad}$}.}
\end{figure}

It is important to note that theoretical predictions for the spin-dependent
parts of $\bar{p}d$  cross section exhibit fairly strong model dependence.
One can compare the predictions for the polarization degree following
from the Nijmegen model with that from the Julich models (see Fig.~\ref{fig:Polarization_d_compare}).
The predictions following from the Julich models are taken from Ref.~\cite{UzikHaiden11}.
Note that they are different from that in Ref.~\cite{UzikHaiden09}.
The polarization degree predicted by the Nijmegen model is about two
or three times larger than that predicted by the Julich models and
transverse polarization degree even has different sign.

\section{Conclusion}

We have calculated the cross section of antiproton-deuteron scattering
making use of the Nijmegen nucleon-antinucleon potential and the Glauber
theory for describing the scattering by a deuteron. Our results show
the possibility to describe total and differential unpolarized $\bar{p}d$
cross section in the whole energy region where the experimental data
exist. The standard Glauber approach turned out to be sufficient for
the precise description of the scattering data. The modifications
to account for the spin dependence of the scattering amplitudes and
the D-wave part of the deuteron wave function are necessary only for
large-angle scattering. The Glauber theory proved to be applicable
for $\bar{p}d$ scattering at rather low energies down to $\unit[50]{MeV}$.

We have also calculated the spin-dependent parts of $\bar{p}d$  cross
section taking shadowing effects into account. Our results indicate
that polarized deuterium target can be used instead of the hydrogen
target with similar or even higher efficiency. However, one can see
fairly strong model dependence of the spin-dependent parts of the
cross section. The Nijmegen model predicts higher polarization degree
than the other models and this was the case for $\bar{p}p$ scattering
too. Only experimental investigation of polarized $\bar{p}p$ or $\bar{p}d$
cross sections can show us what model is closer to the reality.
\begin{acknowledgments}
We are grateful to A.~I.~Milstein for valuable discussions. The
work was supported in part by the Grant 14.740.11.0082 of Federal
Program {}``Personnel of Innovational Russia''.

\newpage{}\end{acknowledgments}

\end{document}